\documentclass[twocolumn,prc,floatfix,superscriptaddress]{revtex4}

\usepackage{color} 
\usepackage{tabularx} 
\usepackage{epsfig}
\usepackage{amsmath} 
\usepackage{amssymb} 
\usepackage{bm}
\usepackage{graphicx}
\usepackage{wasysym}
\usepackage{epsfig}
\usepackage[final]{pdfpages}
\usepackage{xspace}
\usepackage{float}
\usepackage{booktabs}
\usepackage{pgf}
\usepackage[flushleft]{threeparttable}

\newcommand{\gf}{\textsc{Geant4}\xspace}
\newcommand{\mrm}{\textsc{MinRMatrix}\xspace}
\newcommand{\azure}{\textsc{Azure2}\xspace}

\begin{document}

\title{Search for the high-spin members of the $\alpha$:2n:$\alpha$ band in $^{10}$Be}

\author{S. Upadhyayula}
	\email{sriteja@tamu.edu}
\affiliation{Cyclotron Institute, Texas A\&M University,
College Station, Texas 77843, USA}
\affiliation{Department of Physics\&Astronomy, Texas A\&M University,
College Station, Texas 77843, USA}
\author{G. V. Rogachev}
	\email{rogachev@tamu.edu}
\affiliation{Cyclotron Institute, Texas A\&M University,
College Station, Texas 77843, USA}
\affiliation{Department of Physics\&Astronomy, Texas A\&M University,
College Station, Texas 77843, USA}
\affiliation{Nuclear Solutions Institute, Texas A\&M University, College Station, TX 77843, USA}
\author{J. Bishop}
\author{V. Z. Goldberg}
\affiliation{Cyclotron Institute, Texas A\&M University,
College Station, Texas 77843, USA}
\author{J. Hooker}
\altaffiliation[Present address: ]{Department of Physics \& Astronomy, University of Tennessee, Knoxville, TN 37996, USA}
\author{C. Hunt}
\author{H. Jayatissa}
\altaffiliation[Present address: ]{Physics Division, Argonne National Laboratory, Argonne, IL 60439, USA}
\affiliation{Cyclotron Institute, Texas A\&M University,
College Station, Texas 77843, USA}
\affiliation{Department of Physics\&Astronomy, Texas A\&M University,
College Station, Texas 77843, USA}
\author{E. Koshchiy}
\author{E. Uberseder}
\affiliation{Cyclotron Institute, Texas A\&M University,
College Station, Texas 77843, USA}
\author{A. Volya}
\affiliation{Department of Physics, Florida State University, Tallahassee, Florida 32306, USA}
\author{B. T. Roeder}
\author{A. Saastamoinen}
\affiliation{Cyclotron Institute, Texas A\&M University,
College Station, Texas 77843, USA}

%\date{\today}

\begin{abstract}
\begin{description}
\item[Background] Clustering plays an important role in the structure of $^{10}$Be. Exotic molecular-like configurations, such as $\alpha$:2n:$\alpha$, have been suggested at relatively low excitation energies. 
\item[Purpose] To search for the high-spin states that may belong to the molecular-like $\alpha$:2n:$\alpha$ configuration in $^{10}$Be.
\item[Method] Measuring excitation functions for $^{6}$He+$\alpha$ scattering, populating states in the excitation energy range from 4.5 MeV to 8 MeV in $^{10}$Be using a $^6$He rare-isotope beam and a thick helium gas target.
\item[Results] No new excited states in $^{10}$Be have been observed. Stringent limitation on the possible degree of $\alpha$-clustering of the hypothetical yrast 6$^+$ state has been obtained.
\item[Conclusions] The high-spin members of the $\alpha$:2n:$\alpha$ molecular-like rotational band configuration, that is considered to have a 0$^+$ bandhead at 6.18 MeV, either do not exist or have small overlap with the $^{6}$He(g.s.)+$\alpha$ channel.
\end{description}
\end{abstract}

\maketitle

\section{Introduction}

The role of clustering in $^{10}$Be has been a subject of extensive experimental and theoretical scrutiny for the past four decades, since the Molecular Orbital (MO) model was introduced to describe the structure of neutron-rich Be and B isotopes \cite{Okabe1977,Okabe1979,Seya1981}. The dimer $\alpha$ + $\alpha$ structure of $^{10}$Be bound states has been qualitatively discussed by W. von Oertzen in Ref. \cite{Oertzen1997} and confirmed by microscopic antisymmetrized molecular dynamics (AMD) calculations \cite{Kanada1999}, that do not make any {\it a priori} assumptions of clustering. A detailed review paper on the chemical bonding (molecular-like) structures in $^{10}$Be and $^{12}$Be has been published by Ito and Ikeda \cite{Ito2014}. 

It is generally believed that the level structure of $^{10}$Be can be described reasonably well as having the two-center $\alpha$+$\alpha$ structure bonded together by two neutrons that are orbiting the two $\alpha$-particle clusters. The single-particle levels of neutrons in this two-center system are then analogous to the molecular orbitals of electrons in diatomic molecules like H$_2$. The levels in $^{10}$Be can then be assigned to specific molecular orbital configurations. For example, the ground state of $^{10}$Be would correspond to the $({\pi}^-_{3/2})^2$ configuration, in which both neutrons orbit the two-center $\alpha$+$\alpha$ system in the plane perpendicular to the axis that connects the two $\alpha$-particles (see \cite{Ito2014} for a detailed discussion). Of particular interest for this work is the $({\sigma}^+_{1/2})^2$ MO configuration that corresponds to the two neutrons orbiting the two $\alpha$-particles along the $\alpha$+$\alpha$ axis. The inter-cluster distance for this configuration is large due to energy gain associated with the increased radius of ${\sigma}^+_{1/2}$ orbitals. The moment of inertia of the states that belong to this rotational band is therefore large compared to the ground state. Using the more conventional language of shell model, in the limit of zero inter-cluster distance, the $({\sigma}^+_{1/2})^2$ configuration becomes $\nu$(2s1d)$^2$, or (1p)$^4$(2s1d)$^2$ if p-shell  nucleons are included.
\begin{figure}[h]
	\centering
	        \includegraphics[width=\columnwidth]{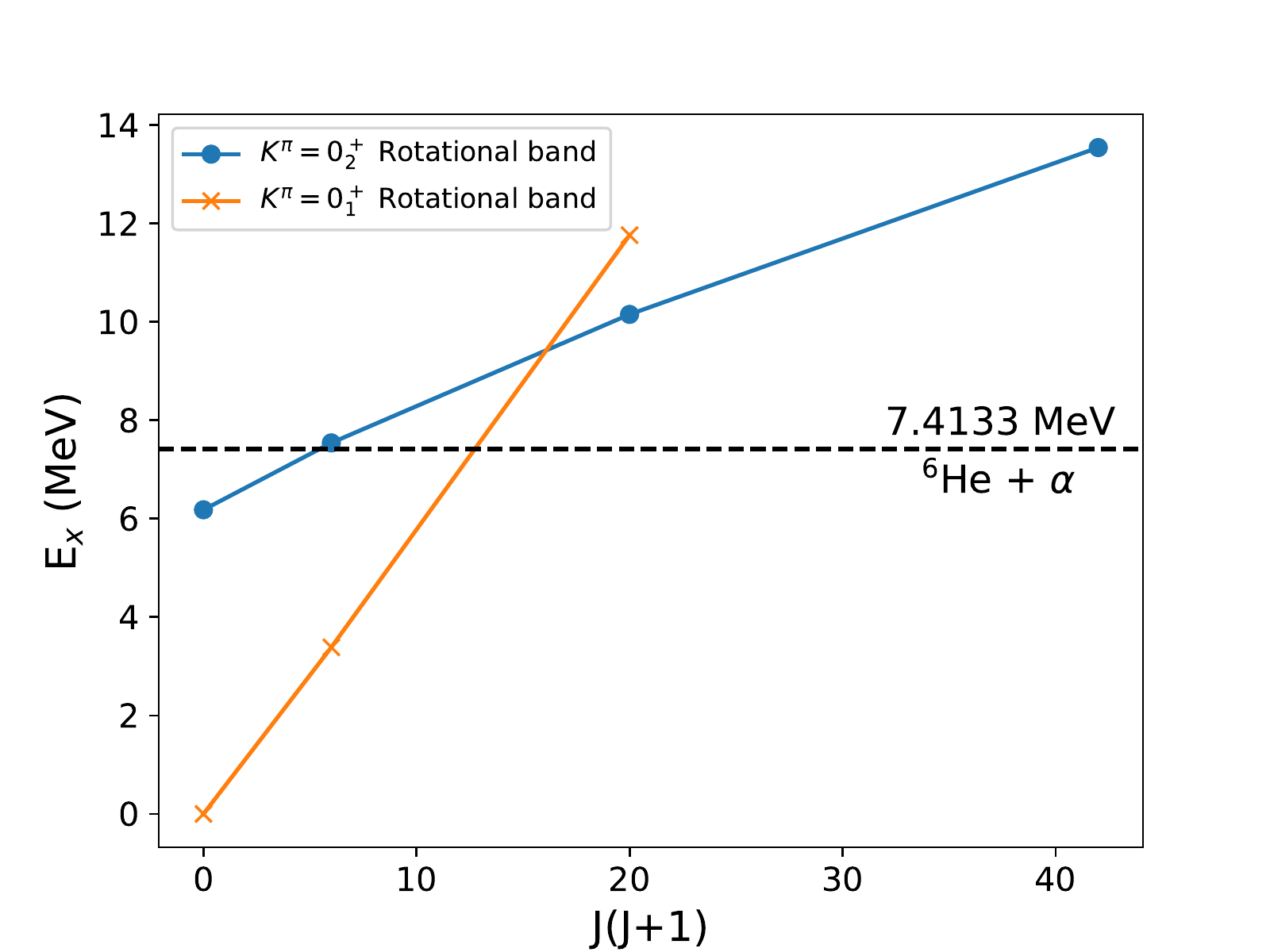}
	\caption{(Color online) The energy-spin systematics for states in the $K^\pi = 0^+_1$ and $K^\pi = 0^+_2$ rotational band. The excitation energies are plotted as a function of the angular momentum $J(J+1)$. }
	\label{fig:angularMomentum}
\end{figure}
There is strong experimental evidence that some states in $^{10}$Be exhibit molecular-like $\alpha$:2n:$\alpha$ configuration \cite{Ito2006}. Theoretically, these exotic structures can be explored microscopically in the antisymmetrized molecular dynamics plus Hartree-Fock approach \cite{Dote1997} or in a Molecular Orbital model \cite{Abe1973}. Based on these theoretical studies, it appears that the 6.179 MeV 0$^+$ state in $^{10}$Be has a pronounced $\alpha$:2n:$\alpha$ configuration with an $\alpha$-$\alpha$ inter-distance of 3.55 fm. This is 1.8 times more than the corresponding value for the $^{10}$Be ground state. The 2$^+$ at 7.542 MeV in $^{10}$Be is believed to be the next member of this rotational band \cite{Nishioka1984}. The state at 10.2 MeV was identified as the next 4$^+$ member \cite{Hamada1994}. While alternative, 3$^-$, spin-parity assignments has been made for this state before \cite{Curtis2001}, we believe that the later experiments \cite{Milin2005,Freer2006} provide a more reliable 4$^+$ spin-parity assignment. Experimental data on $\alpha$-reduced widths \cite{Freer2006} and spectroscopic factors for the 6.179, 7.542, and 10.2 MeV states are consistent with the highly-clustered nature of these states and support assigning them to a single rotational band. Provided that these three states are indeed the members of the same rotational band, the moment of inertia of this band is very large, which supports the molecular-like $\alpha$:2n:$\alpha$ picture for this band. However, the 4$^+$ at 10.2 MeV is not the band-terminating state. The algebraic model based on SU(3) symmetry \cite{Wolsky2010} predicts that this rotational band (with bandhead at 6.179 MeV) should be designated as ($\lambda$,$\mu$)=(8,0) and therefore should contain the 6$^+$ and 8$^+$ states, which are predicted to be the yrast states in $^{10}$Be. Observation of the 6$^+$ state is the main goal of this experiment. Fig. \ref{fig:angularMomentum} shows the energy-spin systematics of the rotational bands built on the ground state as well as the 0$^+$ bandhead at 6.18 MeV.

\begin{figure}[h]
	\centering
		\includegraphics[width=\columnwidth]{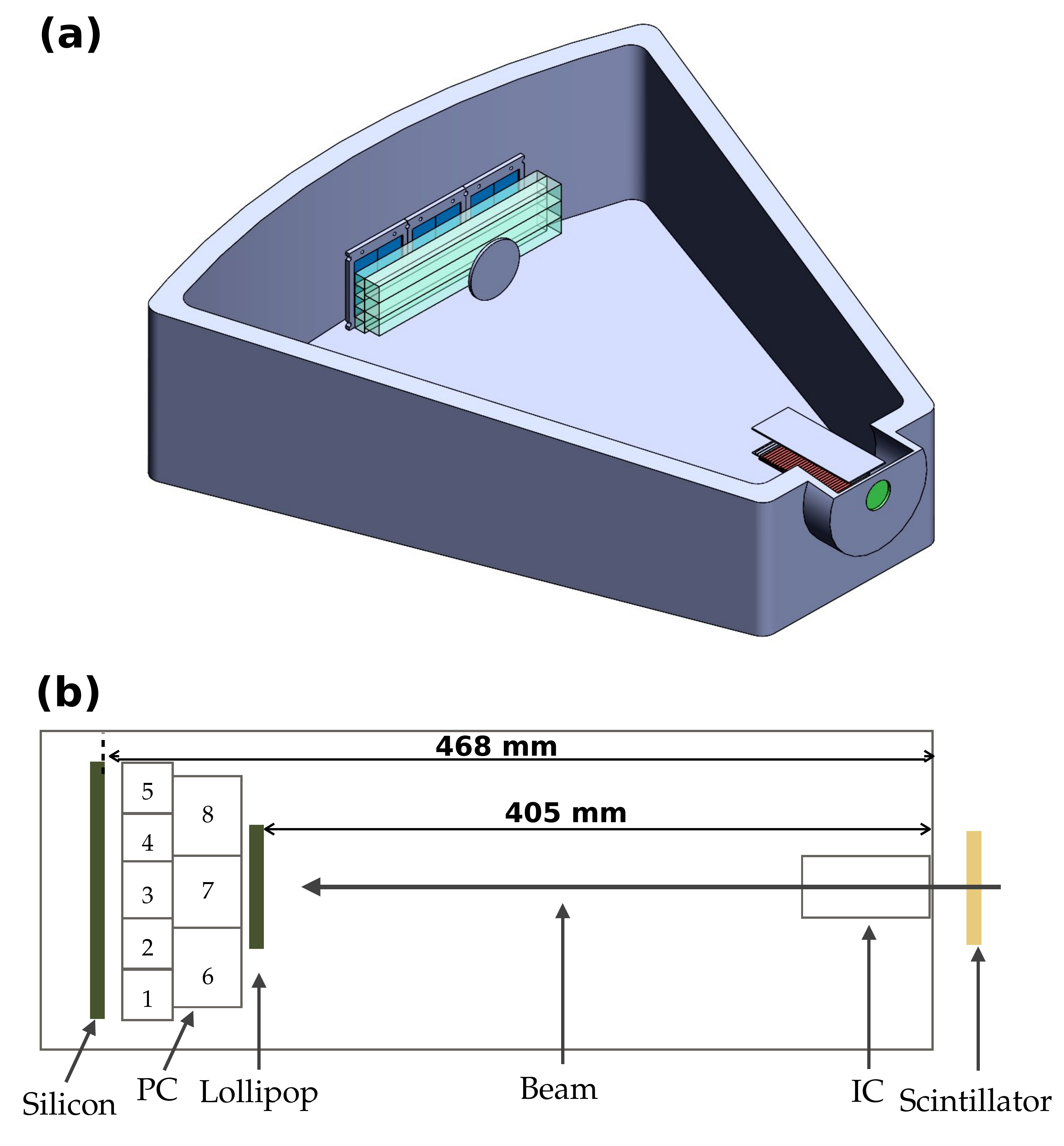}
	\caption{(Color online) (a). CAD rendering of the scattering chamber and detector set up. (b). Cross-sectional view of the set up. The silicon detectors plane, the beam stopper ``Lollipop'', the proportional counter cells (indicated by numbers), the ionization chamber ``IC'' and the scintillator are shown. }
	\label{setup}
\end{figure}

The first indication for a structure that could be interpreted as a relatively narrow high-spin, positive parity state at 13.4 MeV excitation energy in $^{10}$Be were obtained in the excitation function for $^{6}$He+$\alpha$ resonance elastic scattering measured at angles close to 90 degrees in the center of mass (c.m.) \cite{Rogachev2014, Kuchera2013}. Due to a limited measured angular range and the dominance of what appears to be non-resonant elastic scattering, the authors of \cite{Rogachev2014} were not able to make firm conclusions and they made a statement that ``new experimental data at angles close to 180$^{\circ}$ may be necessary to confirm or rule out the existence of this state''. There have been more observations of a state around 13 MeV more recently. It was observed in $^{10}$Be breakup to $\alpha$ and $^6$He in \cite{DellAquila2016} and \cite{Jiang2017}. Statistics were limited in both experiments. The tentative spin-parity assignment was performed in \cite{DellAquila2016} and it was consistent with 6$^+$. The experimental data presented in this paper provide a stringent test on the existence of a highly-clustered 6$^+$ by measuring the excitation function for $^6$He+$\alpha$ elastic scattering at angles where the state should have highest cross section - close to 180$^{\circ}$ in c.m (0$^{\circ}$ in the lab. reference frame).

\section{Experiment}

An experiment was performed to search for the 6$^+$ state in $^{10}$Be at around 13 MeV excitation energy using $^6$He+$\alpha$ scattering. A primary beam of 60 MeV $^7$Li was impinged on a LN$_2$-cooled gas cell filled with D$_2$ at pressure of $1604$ Torr. The $^6$He beam at an energy of 42 MeV was produced in the d($^7$Li,$^6$He)$^3$He reaction and selected by the Momentum Achromat Recoil Separator (MARS) \cite{MARS} at the Cyclotron Institute at Texas A\&M University. The secondary beam had an intensity of $10^4$ particles per second and a purity of $30\%$, with tritium as the main contaminant. Diagnostics of the beam were performed using a $190.5$-$\mu$m thick scintillator monitored by two photomultiplier tubes oriented at 90$^{\circ}$ with respect to the beam axis.  A total of $5.3\times10^{9}$ $^6$He ions were accumulated during the three days run. Another function of the scintillator foil was to degrade the $^6$He beam energy down to 22 MeV, making it more suitable for the measurements in question. The outline of the experimental setup is shown in Fig. \ref{setup}.

The secondary beam entered the scattering chamber, filled with He(96\%)+CO$_2$(4\%) gas mixture at pressure of 1700 Torr, through the 5 $\mu$m entrance (Havar) window. Three forward MSQ25-1000 Micron Semiconductor silicon detectors \cite{micron} were installed 47 cm from the entrance window at forward angles, including zero degrees, to measure the total energy of the recoils. Each silicon detector consisted of four 25 $\times$ 25 mm$^2$ active area independent quadrants. A windowless, resistive readout, position-sensitive wire proportional counter, was installed immediately upstream from the silicon detectors. This consisted of eight cells arranged into two layers, and was used for particle identification and scattering angle reconstruction. A windowless ionization chamber installed at the entrance of the gas-filled scattering chamber was also used in conjunction with the upstream scintillator for overall normalization and beam contaminant identification. The gas pressure was optimized to measure the $\alpha$+$^6$He resonance scattering excitation function between 4.5-8 MeV in c.m. At this pressure of 1700 Torr, the $^6$He ions deposited only 50\% of their total energy in the gas. To avoid saturation of the DAQ trigger rate and damage of the $0^{\circ}$ silicon detector by direct $^6$He beam ions, a removable 2-cm diameter aluminum disk was installed just upstream of the proportional counter cells. It blocked 95\% of the beam. The geometry of the setup was optimized to measure elastic and inelastic $^6$He+$\alpha$ scattering at the lowest laboratory angles possible (closest to 180$^{\circ}$ in c.m.), where the 6$^+$ state is expected to have maximum cross section (which decreases sharply for smaller c.m. angles).

\begin{figure}[ht]
	\centering
		\includegraphics[width=\columnwidth]{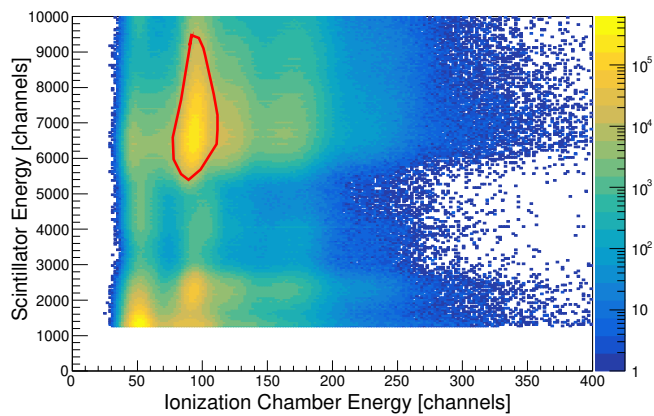}
	\caption{(Color online) Scatter plot of the secondary beam energy deposition in the scintillator foil vs energy deposition in the ionization chamber. The region enclosed within the red contour represents the $^6$He ions. Tritium, the main contaminant in the secondary beam, can be observed at the bottom left.}
	\label{beamSelection}
\end{figure}

 \section{Analysis}
\subsection{Experimental $\alpha$-particle spectrum}

The $^6$He beam ions have an energy of 22 MeV right after the entrance window and 10 MeV just before the beam-stopping aluminum disk, covering a $^6$He+$\alpha$ c.m. energy range from 4 to 9 MeV. The spectrum of $\alpha$-particles that result from $^6$He+$\alpha$ scattering is determined not only by the elastic scattering events, but also by other reactions that produce $\alpha$-particles, such as inelastic scattering and $^6$He breakup. Also, the specific energy losses of $^6$He and $\alpha$-particles are close enough to cause difficulties in particle ID. Therefore, the challenges in the analysis of these experimental data can be classified as follows:

\begin{itemize}
	\item Clean identification of events associated with incoming $^6$He ions.
	\item Particle ID of the detected recoil.
	\item Determining the origin of the measured $\alpha$-particles.
\end{itemize}
  
The first challenge can be addressed by gating on coincidences between a signal in the silicon detector, the scintillator upstream of the scattering chamber and setting a gate on the energy deposited by a beam ion in the scintillator and ionization chamber. Fig. \ref{beamSelection} shows a 2D scatter plot for events that were triggered by a silicon detector; the abscissa is the energy deposited in the ionization chamber and the ordinate is the energy deposited in the scintillator. The events that were produced by $^6$He beam ions were identified using the gate shown in Fig.~\ref{beamSelection}.

\begin{figure}[ht]
	\centering
		\includegraphics[width=\columnwidth]{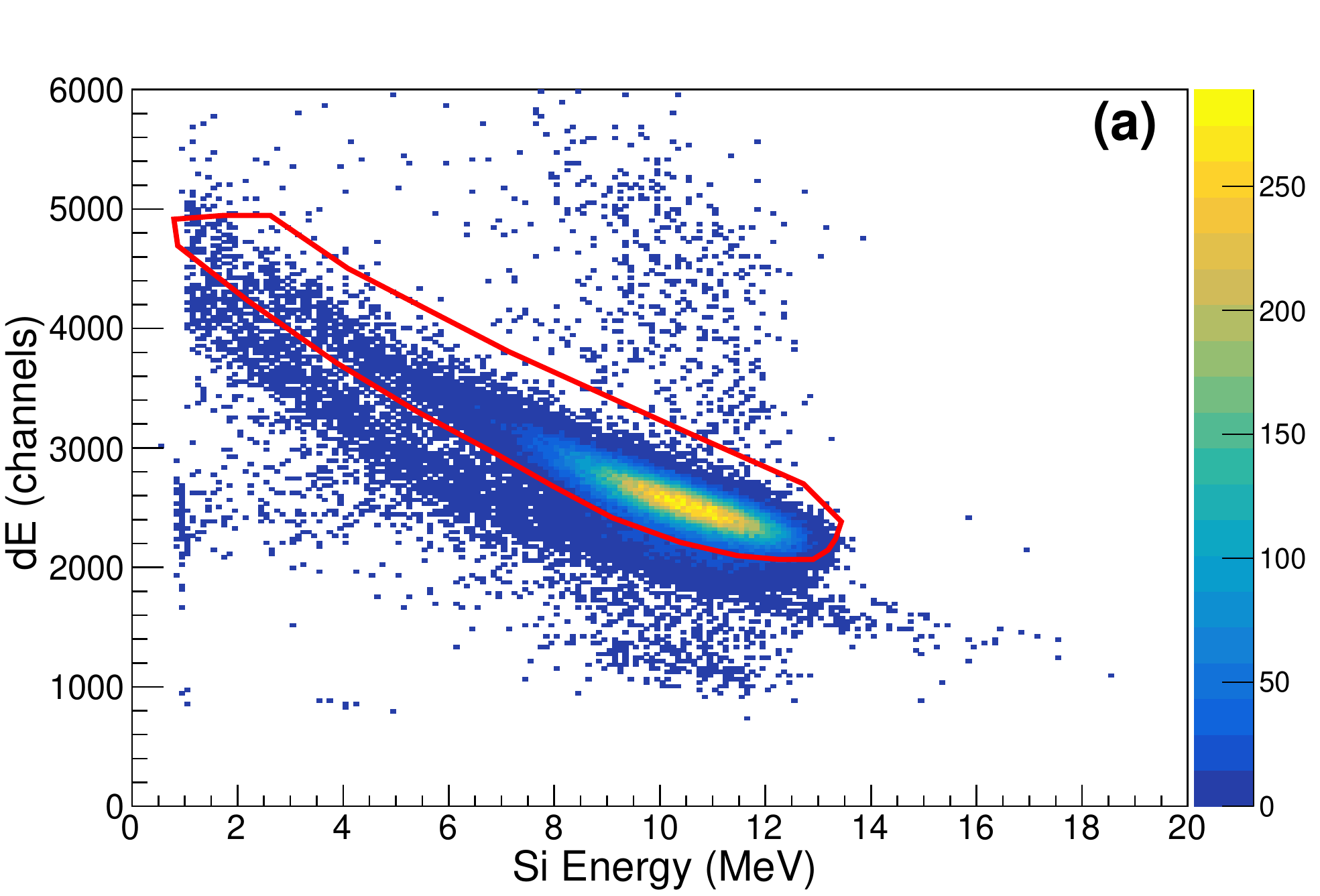}
		\includegraphics[width=\columnwidth]{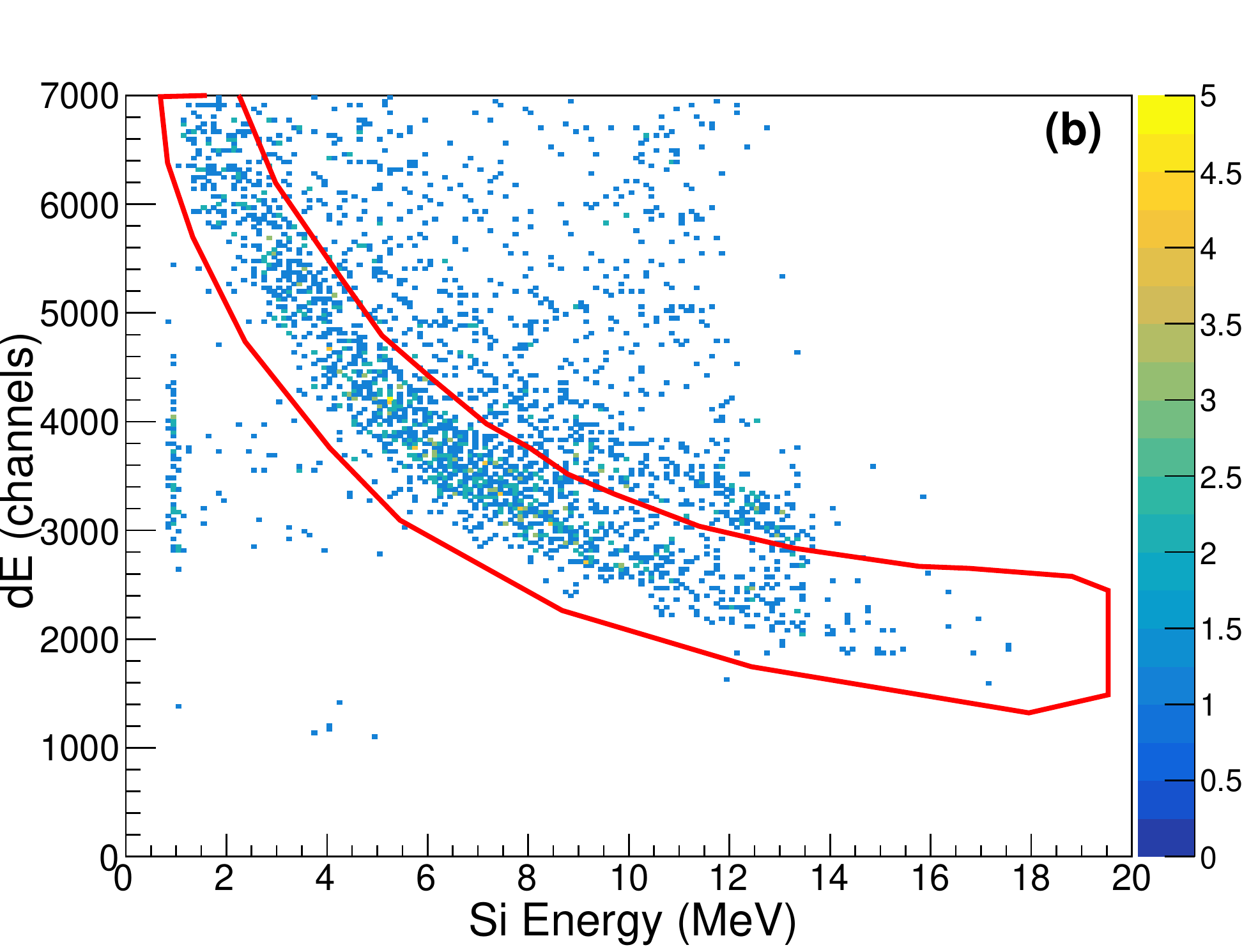}
	\caption{(Color online) Scatter plot of energy losses of the recoils in the proportional counter layers as a function of energy deposited in the Si detector. The first layer (a) of the proportional counter wires is used for an anti-gate (red contour) on the events associated with the $^6$He ions in the proportional counter cell. The anti-gated scatter plot for the second layer of the proportional counter wires is shown in (b). The events selected with the red contour in the second layer (b) are the recoil $\alpha$-particles.}
	\label{alphaSelection}
\end{figure}

To address the second challenge, i.e. the similarity of $^6$He and $\alpha$-particles' specific energy losses and the particle ID issues associated with it, the double-layer proportional counter was used. By performing an anti-gate cut on the $^{6}$He particles in the first layer and gating on the $\alpha$-particles in the second layer, a clean $\alpha$-particle spectrum can be extracted from the data. The corresponding sequence is shown in Fig. \ref{alphaSelection}. By gating just above and below the $\alpha$-band in the second layer, we estimate that no more than 10\% of counts in the $\alpha$-spectrum correspond to the misidentified $^6$He and virtually all of them are located around the $^6$He beam energy in the silicon detector between 10 and 12 MeV. 

\begin{figure}[h]
	\centering
	\includegraphics[width=\columnwidth]{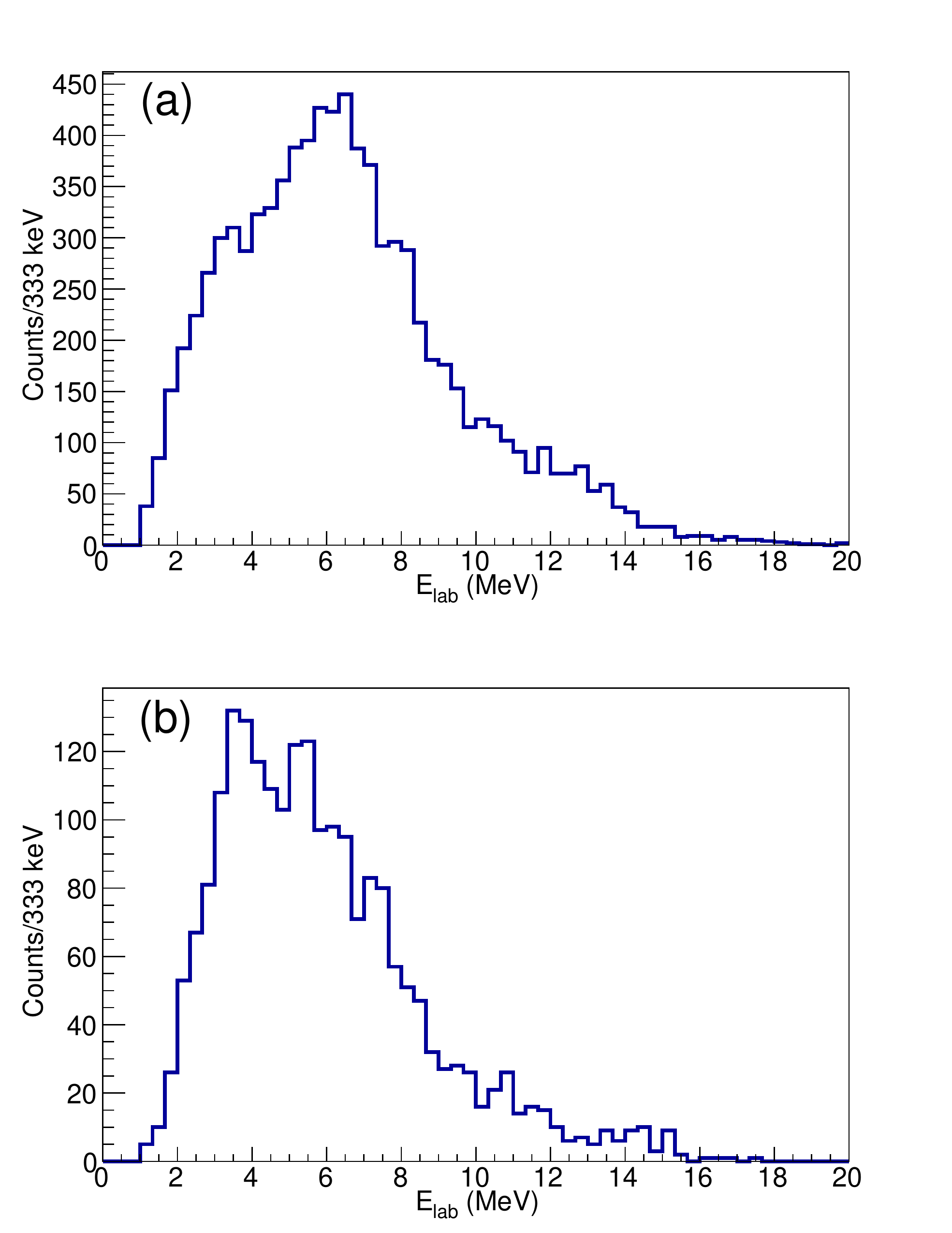}
	\caption{(Color online) $\alpha$-particle spectra for regions (a) and (b) of the silicon detectors. Region (a) corresponds to the inside quadrants of the outside two detectors, and region (b) corresponds to the outside quadrants of the outside detectors.}
	\label{RawSpectra}
\end{figure}

The spectra of $\alpha$-particles for two different angular regions are shown in  Fig. \ref{RawSpectra}. The region (a) corresponds to the inside, and the region (b) corresponds to the outside quadrants of the two outside detectors. For an $^6$He+$\alpha$ c.m. energy of 6 MeV, regions (a) and (b) cover 120$^{\circ}$-170$^{\circ}$ and 100$^{\circ}$-150$^{\circ}$ c.m. scattering angles respectively. The origin of $\alpha$-particles, i.e. the specific reaction process that produced them, could not be determined on an event-by-event basis in these measurements. The obvious feature of the $\alpha$-spectrum in Fig. \ref{RawSpectra} is a peak with a maximum at 7 MeV in region (a) and a lower energy in region (b) (smaller c.m. but larger lab. scattering angles). In principle, this peak may potentially be due to a resonance in the excitation function for $^6$He+$\alpha$ elastic and/or inelastic scattering. Below, we test if the hypothesis that there is a 6$^+$ resonance at 13.5 MeV excitation energy in $^{10}$Be (6.1 MeV in c.m.) is consistent with the experimental data.

\subsection{Hypothesis of strong $\alpha$-cluster 6$^+$ at 13.5 MeV in $^{10}$Be}

To evaluate the sensitivity of these measurements to a hypothetical $6^+$ state at 6.1 MeV in c.m. (13.5 MeV excitation in $^{10}$Be \cite{Rogachev2014, DellAquila2016, Jiang2017}), we used partial widths suggested in Ref. \cite{Kuchera2013} (see Table \ref{table1}). Note that using these parameters for the 6$^+$ state, one can reproduce the excitation function for the $^6$He+$\alpha$ resonance elastic scattering at 90$^\circ$ (Refs. \cite{Rogachev2014, Kuchera2013}). R-matrix calculations were performed using codes \mrm \cite{Johnson2008} and cross-checked with \azure \cite{Azuma2010AZURE}.

\begin{figure}[h]
	\centering
	\includegraphics[width=\columnwidth]{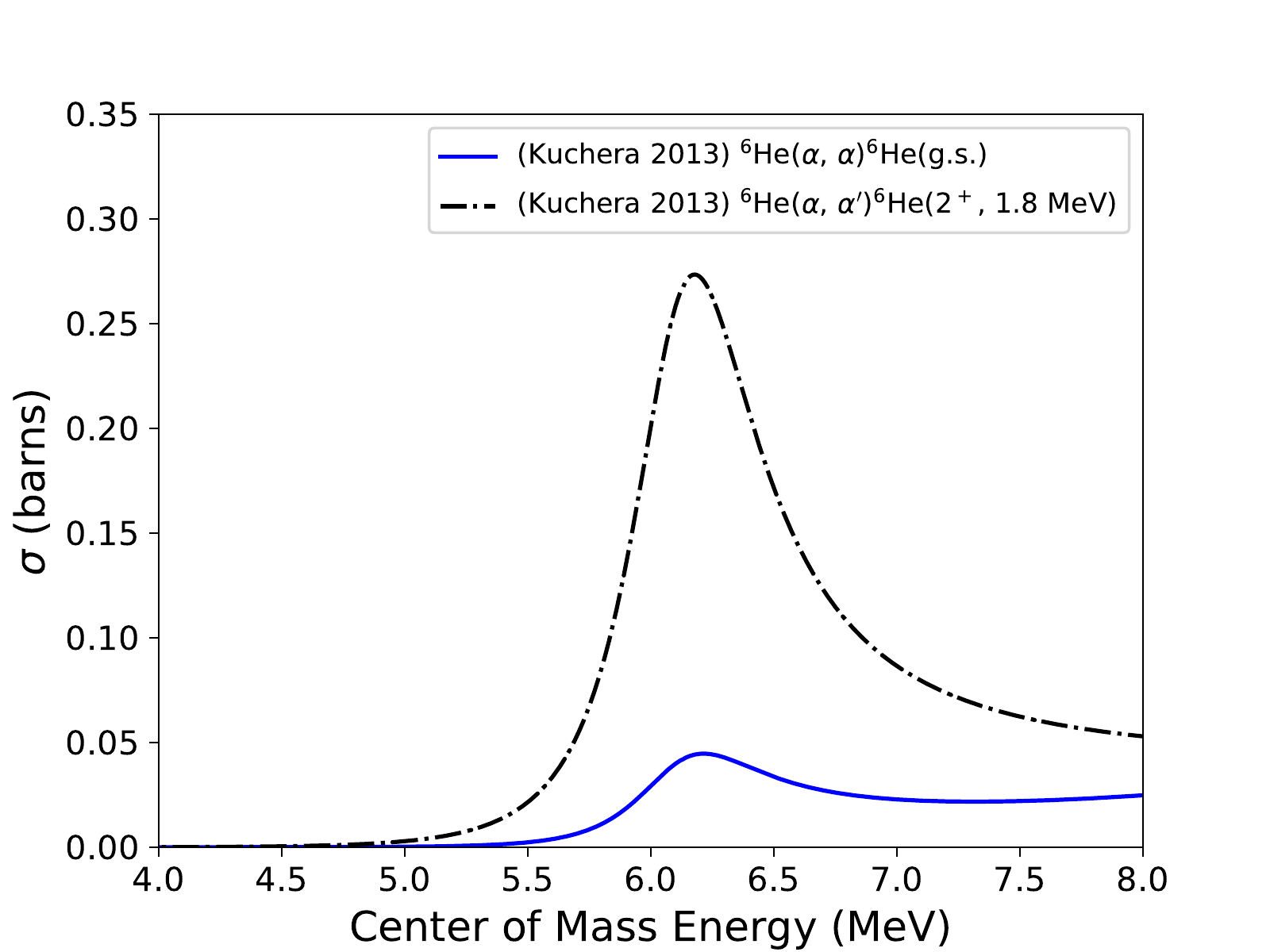}
	\caption{(Color online) Total cross section for inelastic scattering (dash-dotted curve) and 90-180$^{\circ}$ c.m. angle integrated cross section for elastic scattering (solid curve) produced by \mrm calculations using the partial widths for the hypothetical 6$^+$ state from Ref. \cite{Kuchera2013}.}
	\label{fig:kucheraCS}
\end{figure}

The total cross sections for the $^6$He($\alpha$,$\alpha$)$^6$He(2$^+$,1.8 MeV) inelastic scattering and the 90$^{\circ}$ to 180$^{\circ}$ in c.m. angle-integrated cross section for the $^6$He+$\alpha$ elastic scattering are shown in Fig. \ref{fig:kucheraCS}. Both cross sections include the 6$^+$ resonance only.

\begin{figure}[h]
	\centering
	\includegraphics[width=\columnwidth]{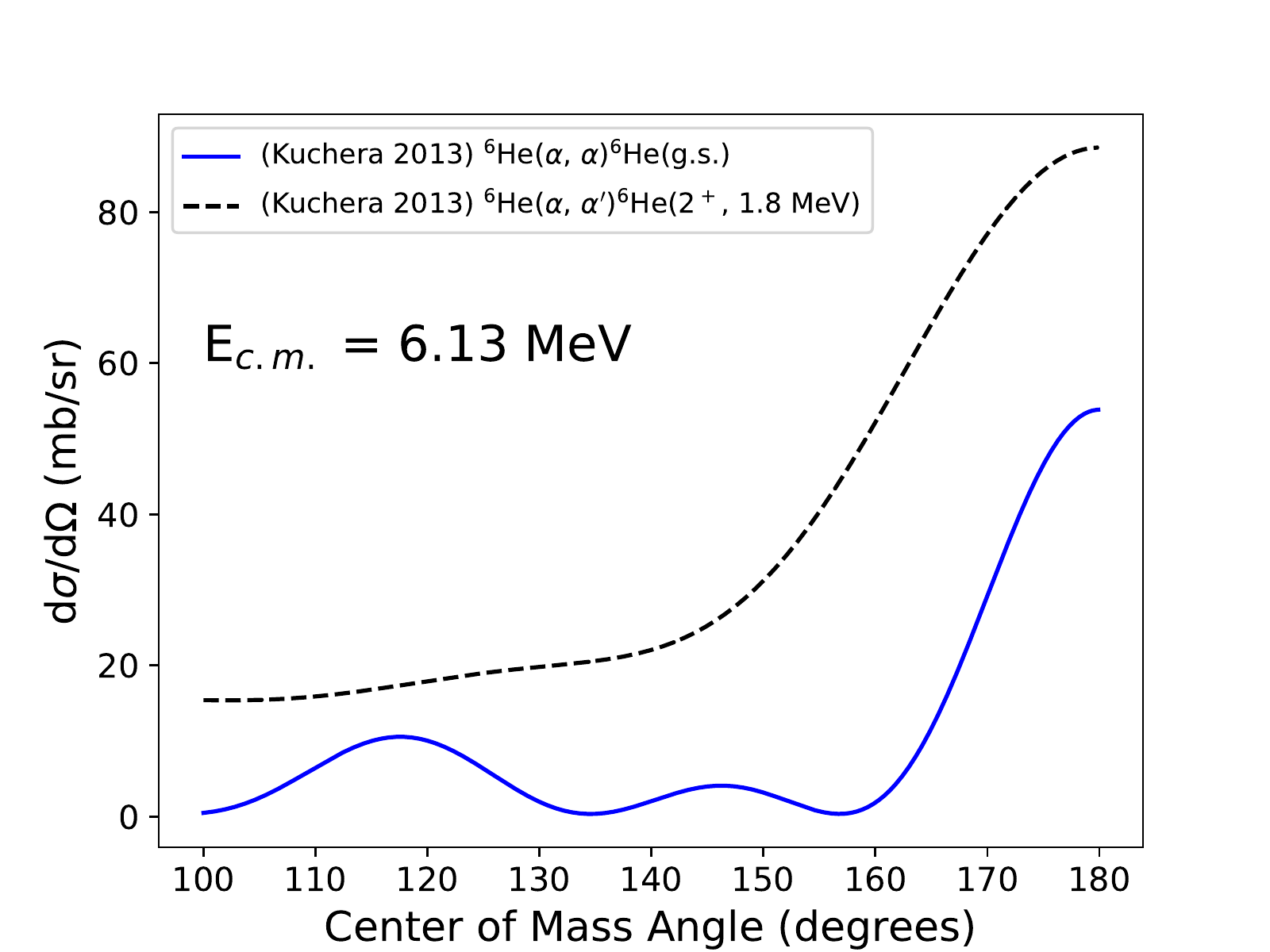}
	\caption{(Color online) Angular distribution for inelastic scattering (dash curve) and elastic scattering (soild curve) produced by \mrm calculations using the partial widths for the hypothetical 6$^+$ state from Ref. \cite{Kuchera2013}. This plot shows a prominent feature at angles close to 180$^{\circ}_{c.m.}$.}
	\label{fig:kucheraAngDist}
\end{figure}

At angles close to 180$^{\circ}$, the 6$^+$ state is a prominent feature (see Fig. \ref{fig:kucheraAngDist}). Differential cross sections as a function of energy and angle from these calculations (Figs. \ref{fig:kucheraCS} and \ref{fig:kucheraAngDist}) were then used to generate the interactions in the \gf -based Monte Carlo simulations. The shape of the $\alpha$-particle spectrum, as well as the total expected yield due to the elastic and inelastic channels of the hypothetical 6$^+$ state, were then obtained. The $\alpha$-particles that result from the decay of the first excited state of $^6$He, populated in the inelastic scattering, were also included in the simulations.

\begin{table*}[ht]
	\caption{Resonance parameters for the yrast 6$^+$ state in $^{10}$Be from \cite{Kuchera2013, DellAquila2016, Jiang2017, Kravvaris2018}, and this work. $\theta^{2}_{\alpha}$ and $\theta^{2}_{\alpha'}$ were calculated using a channel radius of 4.77 fm ($r_0 = 1.4$ fm). The parameters from \cite{Kuchera2013} were determined from a R-Matrix fit. 
	%The parameters for this work were determined by a combination of R-Matrix calculations and \gf simulations. 
	}
	\label{table1}
	\centering
	\begin{ruledtabular}
	\begin{tabular}[t]{c  c  c  c  c  c  c  c  c  c}
%	J$^\pi$ & E$_{c.m.}$ [MeV] & E$_{x}$ [MeV] & $\Gamma_{tot}$ [keV] & $\Gamma_{\alpha}$ [keV] & $\Gamma_{in.}$ [keV] & $\Gamma_{n}$ [keV] & $\theta^{2}_{\alpha}$ & $\theta'^{2}_{\alpha}$ & Ref.\\
%	E$_{c.m.}$ [MeV] & E$_{x}$ [MeV] & $\Gamma_{tot}$ [keV] & $\Gamma_{\alpha}$ [keV] & $\Gamma_{\alpha'}$ [keV] & $\Gamma_{n}$ [keV] & $\theta^{2}_{\alpha}$ & $\theta^{2}_{\alpha'}$ & $\Gamma_{\alpha}$ /$\Gamma_{\alpha'}$ & Ref.\\
%	\hline
%%	$4^+$ & 2.79 & 10.20 & 360 & 185 & 51 & 124 & 1.56  &  0.08  &  \cite{Kuchera2013}\\
%	6.13 & 13.54 & 914 & 99 & 763 & 52 & 0.99  &  1.25  & 0.79 &  \cite{Kuchera2013}\\
%	- & (13.5) & ($<350$) & - & - & - & -  &  -  &  - & \cite{DellAquila2016}\\
%	- & (13.5) & - & - & - & - & -  &  -  &  - & \cite{Jiang2017}\\
%	- & 13.5 & - & - & - & - & 0.1  & 0.66 & 0.02 & \cite{Kravvaris2018}\\
%    6.13 & 13.5 & - & - & - & - & -  & - & (0.01) & This work\\

	E$_{x}$ [MeV] & $\Gamma_{tot}$ [keV] & $\Gamma_{\alpha}$ [keV] & $\Gamma_{\alpha'}$ [keV] & $\Gamma_{n}$ [keV] & $\theta^{2}_{\alpha}$ & $\theta^{2}_{\alpha'}$ & $\Gamma_{\alpha}$ /$\Gamma_{\alpha'}$ & Ref.\\
	\hline
%	$4^+$ & 2.79 & 10.20 & 360 & 185 & 51 & 124 & 1.56  &  0.08  &  \cite{Kuchera2013}\\
	13.54 & 914 & 99 & 763 & 52 & 0.99  &  1.25  & 0.13 &  \cite{Kuchera2013}\\
	(13.5) & ($<350$) & - & - & - & -  &  -  &  - & \cite{DellAquila2016}\\
	(13.5) & - & - & - & - & -  &  -  &  - & \cite{Jiang2017}\\
	- & - & - & - & - & 0.1  & 0.66 & 0.02 & \cite{Kravvaris2018}\\
    13.5 & - & - & - & - & -  & - & $<$0.017 & This work\\
	
	\end{tabular}
	\end{ruledtabular}
	\begin{tablenotes}
		\small
		\item $\Gamma$ = 2 $\theta^2$ $\frac{{\hbar}^2}{\mu a^2}$ $P_l$ and $\theta^2 = \gamma^2/\gamma_w^2$.  $\theta^2$ is the dimensionless reduced width. $\mu$ is the reduced mass and $a$ is the channel radius, $\gamma$ is the reduced width amplitude and $P_l$ is the penetrability factor. 
	\end{tablenotes}
\end{table*}

\begin{figure}[h]
	\centering
	\includegraphics[width=\columnwidth]{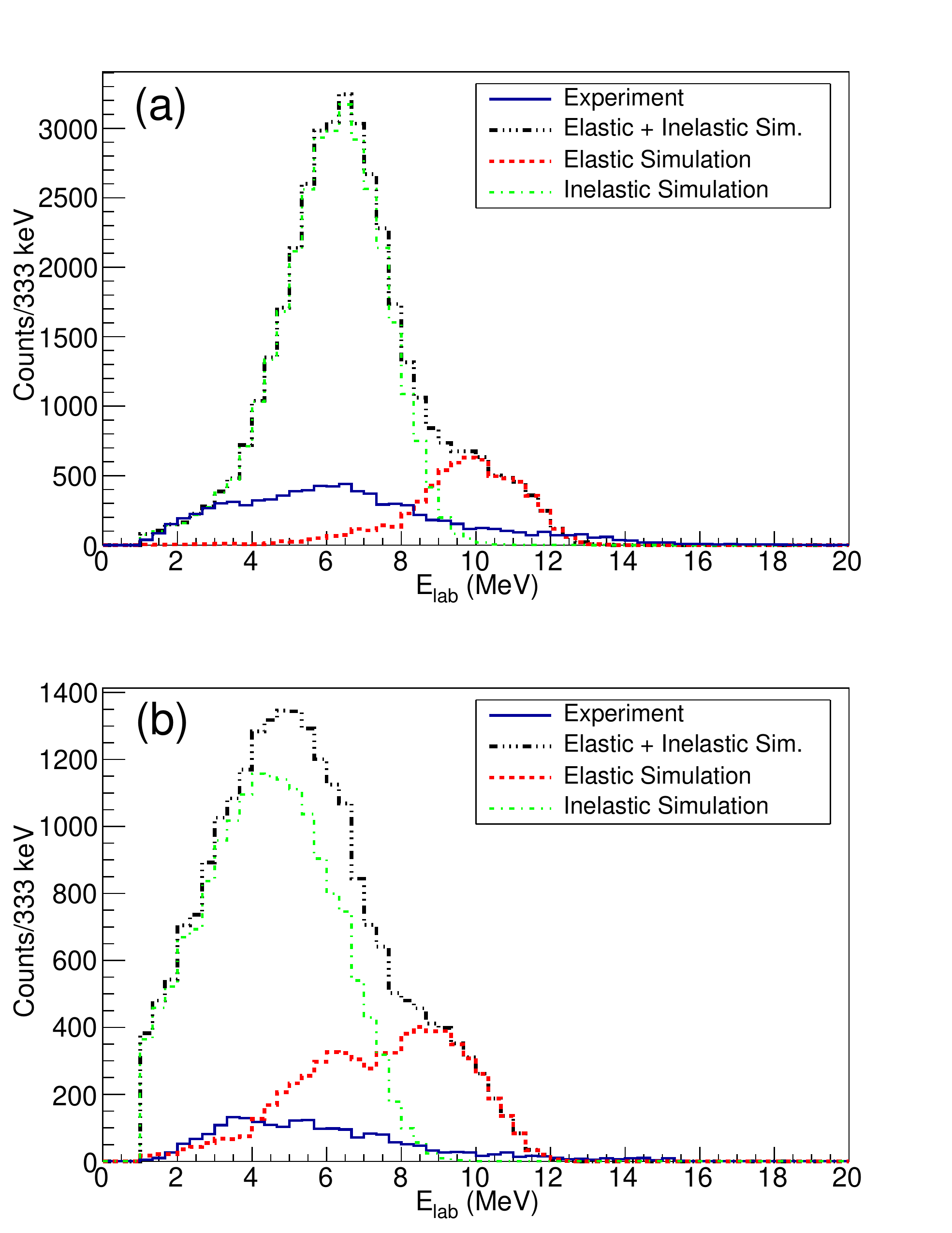}
	\caption{(Color online) \gf Monte Carlo simulations with parameters from Ref. \cite{Kuchera2013} for the elastic (dashed red) and inelastic channels (dashed green), overlaid with this work's experimental spectrum (solid blue) shown in Fig. \ref{RawSpectra}. The sum of the elastic and inelastic spectrum is represented by the dashed black curve. (a) Experimental, elastic, inelastic, and total spectra for region (a). (b) Experimental, elastic, inelastic and total spectra for region (b).}
	\label{fig:kucheraGeant}
\end{figure}

The results from these \gf simulations are shown in Fig. \ref{fig:kucheraGeant}. It is obvious that the event yield produced through the elastic and inelastic scattering in the \gf simulation far exceeds what was observed during our experiment in both regions. From this, we conclude that either there is no 6$^+$ state near 13 MeV excitation energy in $^{10}$Be, or its $^6$He(g.s.)+$\alpha$ partial width is smaller than in Ref. \cite{Kuchera2013}.

\subsection{Hypothesis of a 6$^+$ state with parameters from microscopic calculations \cite{Kravvaris2018}}
\begin{figure}[h]
	\centering
	\includegraphics[width=\columnwidth]{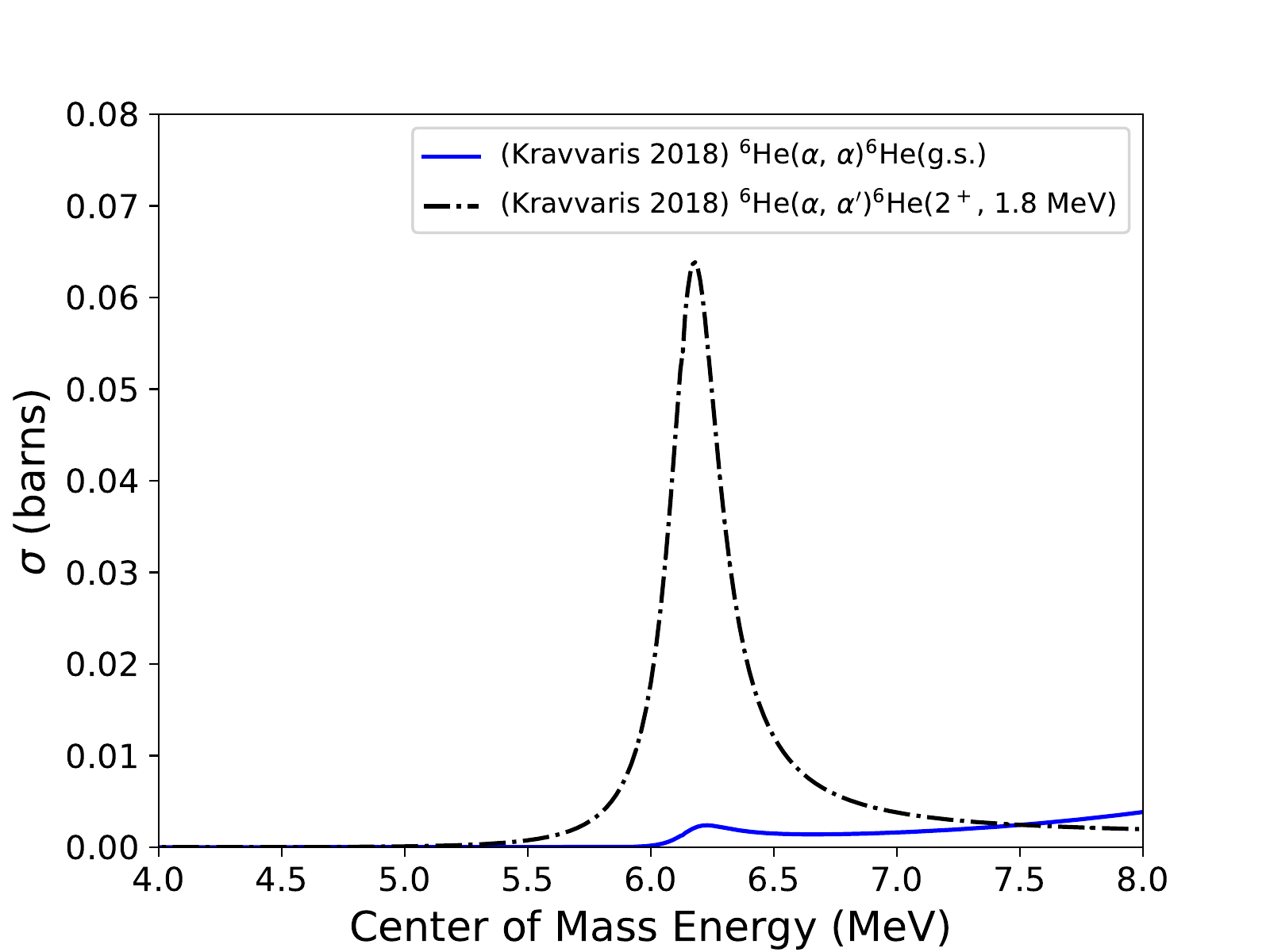}
	\caption{(Color online) \mrm calculations using the spectroscopic factors provided by \cite{Kravvaris2018} to give the total cross sections of the elastic and inelastic channels. In the case of the elastic channel, forward angles were excluded from the calculations to omit the Rutherford contribution. These cross sections were fed in to the \gf simulations, resulting in Fig. \ref{fig:volyaGeant}}.
	\label{fig:volyaCS}
\end{figure}

\begin{figure}[h]
	\centering
	\includegraphics[width=\columnwidth]{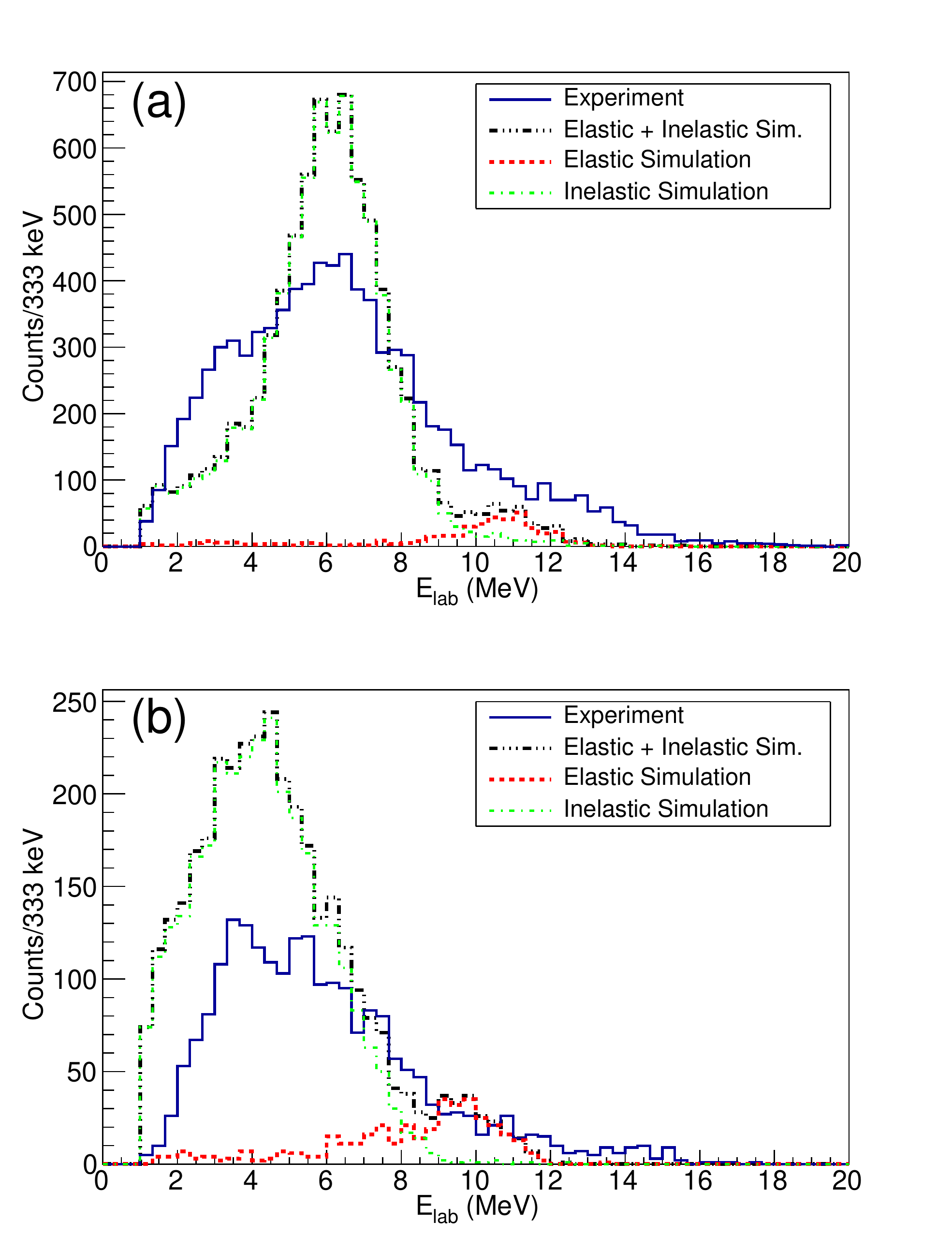}
	\caption{(Color online) \gf Monte Carlo simulations with parameters from Ref. \cite{Kravvaris2018}, overlayed with this work's experimental spectrum shown in Fig. \ref{RawSpectra}. While the \gf simulations with parameters provided by Ref. \cite{Kravvaris2018} still far exceed the expected yield in this work's experimental spectra as shown here, this setup is still sensitive to the elastic and inelastic channel of the $6^+$ state. (a) Experimental (solid blue), elastic (dashed red), inelastic( dashed green), and total spectra (dashed black) for region (a). (b) Experimental (solid blue), elastic (dashed red), inelastic( dashed green) and total spectra (dashed black) for region (b).}
	\label{fig:volyaGeant}
\end{figure}

Resonating Group Method (RGM) and no-core shell model (NCSM) approach, described in detail in Ref. \cite{Kravvaris2019}, was used in \cite{Kravvaris2018} to calculate the $\alpha$ spectroscopic factors for the states in $^{10}$Be. The $\alpha$ spectroscopic factors for the yrast $6^+$ state in $^{10}$Be, that appears at 13.5 MeV in these calculations, are $0.1$ and $0.66$ for the elastic and inelastic channels respectively. The cross sections were calculated for these parameters using \mrm, and subsequently used by \gf to generate realistic spectra that can be compared with our experimental spectrum, in the same way as in the previous section. Consistent with the previous section, angle-integrated cross section for the elastic scattering and total cross section for the inelastic scattering are shown in Fig. \ref{fig:volyaCS}. The resulting spectra from the \gf simulations are shown in Fig. \ref{fig:volyaGeant} and compared to the experimental data.

A $6^+$ state with the spectroscopic factors given in Ref. \cite{Kravvaris2018} would produce a significantly higher event yield at some energies and an $\alpha$-spectraum that has a different shape than the experimentally observed one. Therefore this hypothesis is still not consistent with the experimental data, but it is certainly a significant improvement over the hypothesis discussed in the previous section.

\subsection{Hypothesis of energy independent cross section}

We have thus far compared our data to simulations that assumed existence of the hypothetical 6$^+$ state with parameters from \cite{Kuchera2013} and \cite{Kravvaris2018} and failed to explain the shape and yield of the experimental spectra. Now we assume that the 6$^+$ state does not exist (or that its $^6$He(g.s.)+$\alpha$ partial width is negligible) and simulate energy and angle independent elastic and inelastic cross sections for regions (a) and (b) of the Si array. The absolute magnitudes of the average differential cross sections were varied for the two regions independently.

\begin{figure}[h]
	\centering
		\includegraphics[width = \columnwidth]{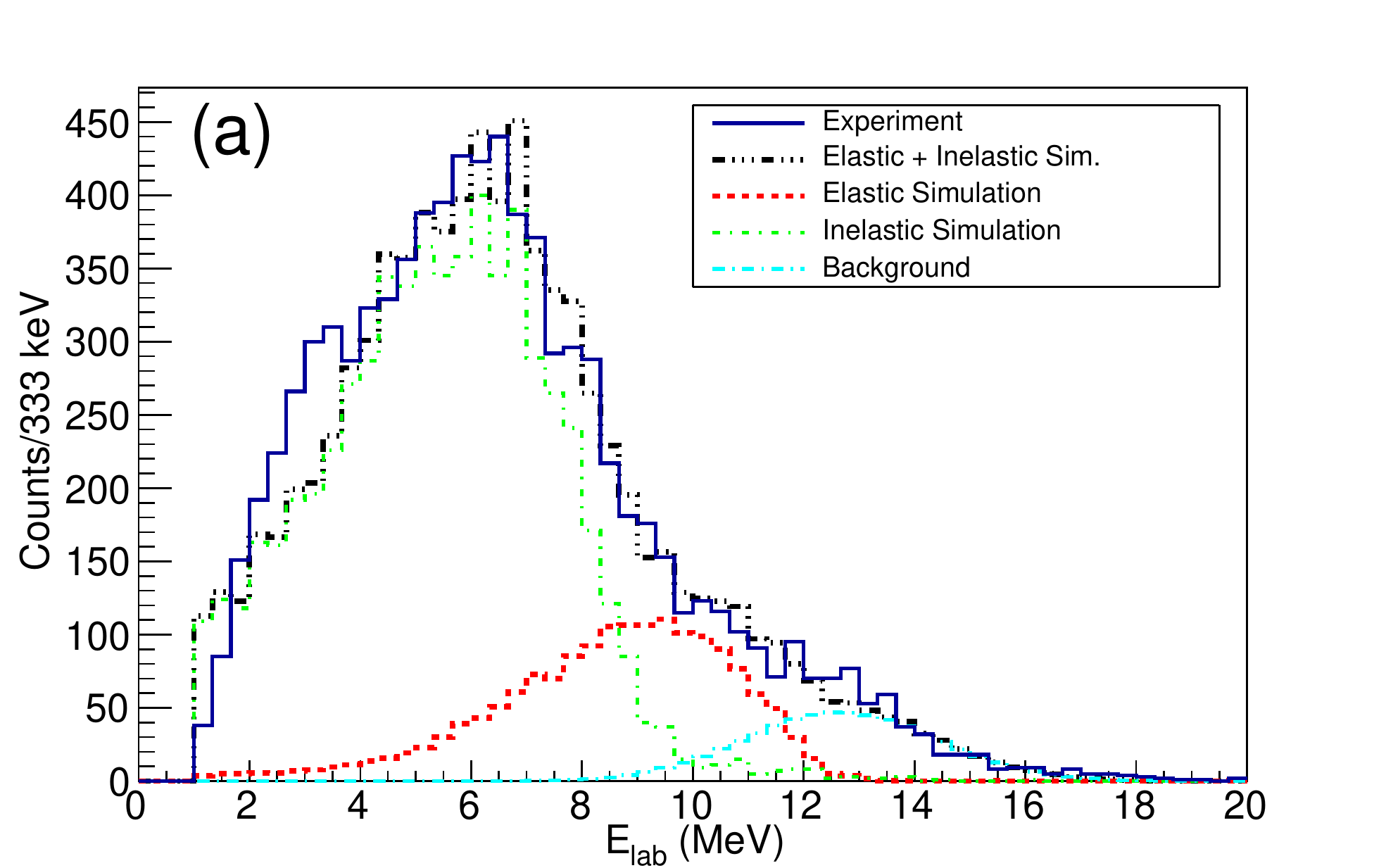}
		\includegraphics[width = \columnwidth]{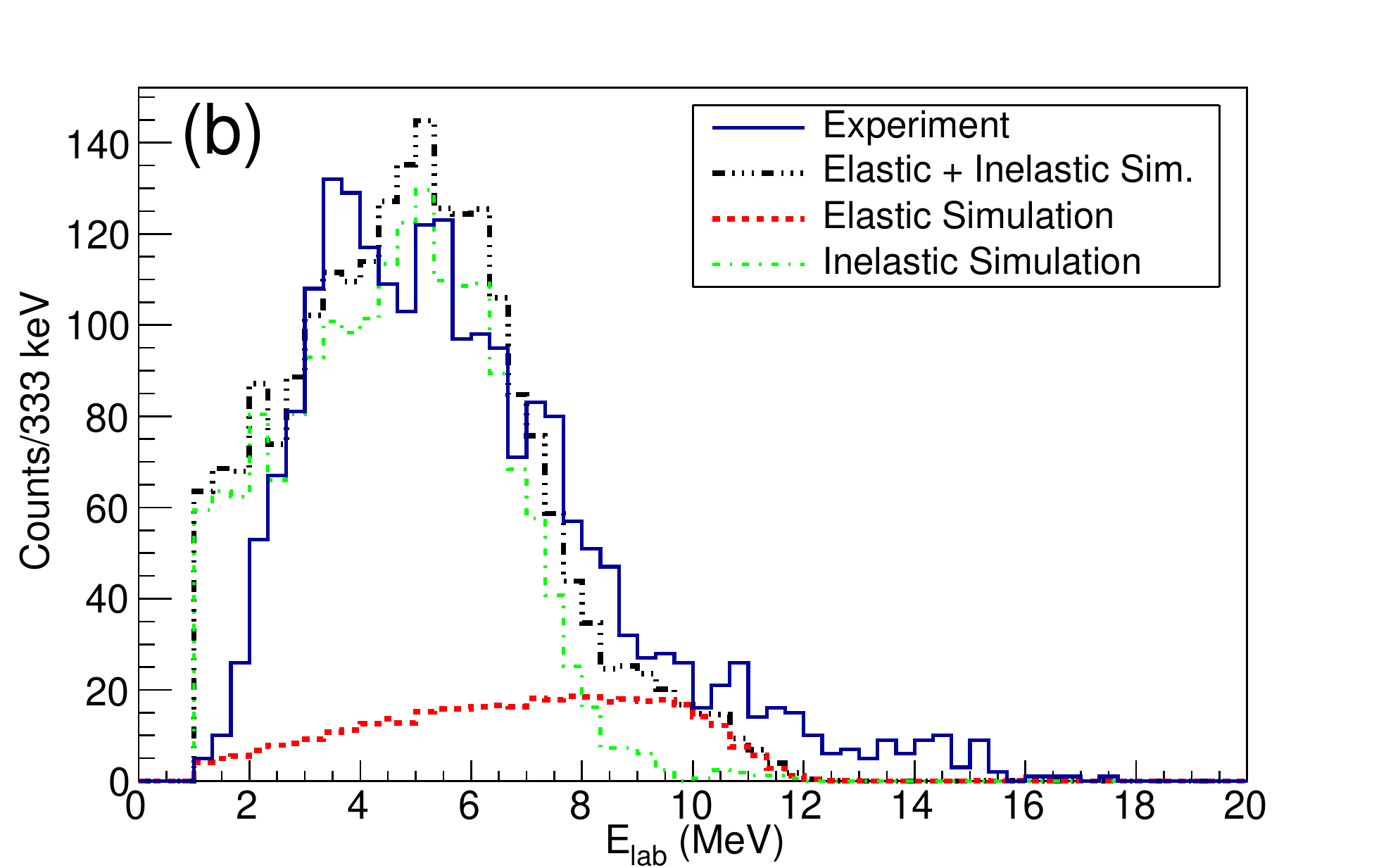}
	\caption{(Color online) \gf Monte Carlo simulations overlaid with this work's experimental spectra. The parameters used for the \gf simulations are consistent with the absence of the $6^+$ resonance. (a) Experimental (solid blue), elastic (dashed red), inelastic (dashed green), and total spectra (dashed black) for region (a). (b) Experimental (solid blue), elastic (dashed red), inelastic (dashed green) and total spectra (dashed black) for region (b).}
	\label{fig:noRes}
\end{figure}

The \gf simulations similar to those described in previous sections were performed. However, we now assume that the cross section does not depend on energy within the relevant excitation energy region - from 4 to 10 MeV in c.m. An almost perfect fit to the observed experimental spectrum (see Fig. \ref{fig:noRes}) can be achieved with the following simple assumptions:

\begin{itemize}
	\item The average differential cross section for the $^6$He($\alpha$,$\alpha$')$^6$He(2$^+$,1.8 MeV) reaction is 1.8 mb/sr for the region (a) and  1.1 mb/sr for region (b).
	\item The average differential cross section for elastic scattering is 0.7 mb/sr for region (a) and 0.2 mb/sr for region (b).
	\item There is a small background that has a shape of Maxwell-Boltzmann distribution at higher energies in region (a) that is unrelated to elastic or inelastic channels accounted for in the previous two assumptions.
\end{itemize}

Note that general trend is consistent with the data presented in Ref. \cite{Suzuki2013}. Specifically, as in Ref. \cite{Suzuki2013}, we observe that the elastic scattering cross section is smaller than the inelastic and that no sharp structures in either elastic or inelastic cross sections are necessary to reproduce the data.  The average cross sections are lower in this work as compared to Ref. \cite{Suzuki2013}. However, direct comparison is challenging because our measurement a covers higher energy range (4.5-8 MeV vs 2-6 MeV in \cite{Suzuki2013}) and different angular range (100$^{\circ}$-170$^{\circ}$ vs 40$^{\circ}$-120$^{\circ}$ in \cite{Suzuki2013}). 

$\alpha$-particles produced as a result of the $^6$He($\alpha$,2n)$^8$Be reaction, the total cross section for which was established in Ref. \cite{Suzuki2013}, have energies that are too low to reach the Si detectors in our setup due to energy losses in the gas. Effectively, our measurement is not sensitive to this reaction.

\subsection{Establishing an upper limit for the partial width of the hypothetical 6$^+$ state}

\begin{figure}[h]
	\centering
		\includegraphics[width = \columnwidth]{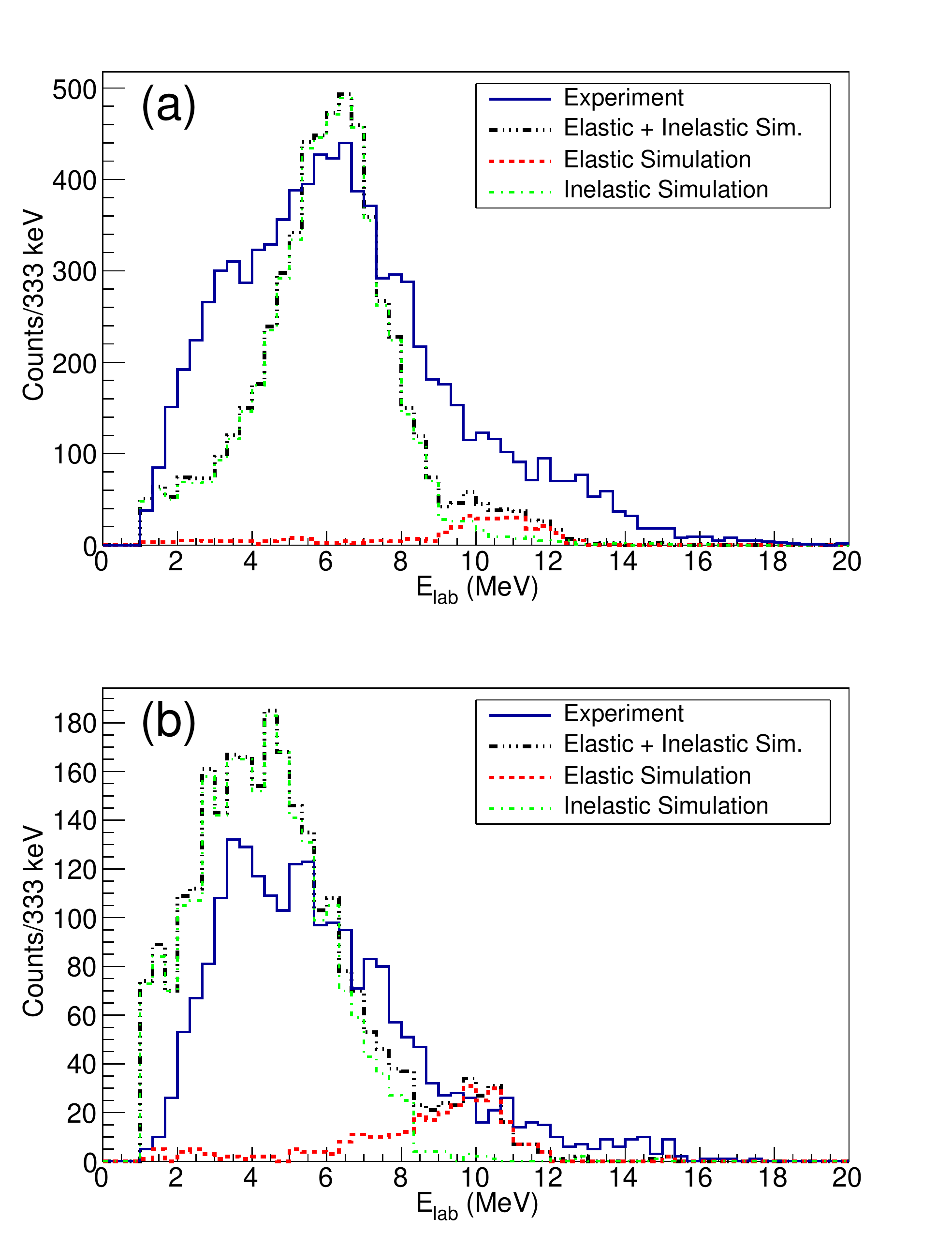}
	\caption{(Color online) \gf Monte Carlo simulations overlaid with this work's experimental spectra. The parameters used for the \gf simulations are $\Gamma_{\alpha}/\Gamma_{\alpha'}$ = 0.017. This plot shows the experimental (solid blue), elastic (dashed red), inelastic (dashed green), and total spectra (dashed black) for region (a).}
	\label{fig:limit}
\end{figure}

The spectroscopic factors from \cite{Kravvaris2018} were shown to produce higher yields than those observed in our experimental spectra. Further analysis shows that due to the fact that the cross section for inelastic scattering (which provides the dominant contribution under all of the scenarios considered above) is proportional to the ratio of $\Gamma_{\alpha}$ to $\Gamma_{\alpha'}$, we can place an upper limit on this parameter. Fig. \ref{fig:limit} shows a simulated spectrum with $\Gamma_{\alpha}/\Gamma_{\alpha'}$ = 0.017 for the 6$^+$ state in comparison with the experimental data. It is evident that the event yield already exceeds the experimental one at this ratio and since no background was introduced in these simulations the $\Gamma_{\alpha}/\Gamma_{\alpha'}$ = 0.017 should be considered a conservative upper limit. Note that it is not too far off from the prediction of the microscopic model of Ref. \cite{Kravvaris2018}, in which the same ratio is 0.02 (see Table \ref{table1}).

A peak at 13.5 MeV in excitation energy of $^{10}$Be in the $^6$He(g.s.)+$\alpha$ coincidence spectrum was observed in the breakup of $^{10}$Be on a CH$_2$ target in Ref. \cite{DellAquila2016}. Tentative spin-parity assignment of 6$^+$ was made for this state in the same work.  It is not surprising that the statistics were rather small in that experiment - the branching ratio for the decay that was used to identify this state is below 1.7\%. It appears that conclusive identification of the 6$^+$ state in $^{10}$Be at 13.5 MeV will require another experiment in which the $\alpha$-decay of this state to the first excited state in $^6$He is measured.
 
\section{Summary}

In summary, we have performed a search for the $6^+$ state at 13.5 MeV in $^{10}$Be in the excitation function for $^6$He + $\alpha$ scattering. This state was suggested as the next member of the molecular $\alpha$:2n:$\alpha$ rotational band in Ref. \cite{Wolsky2010}. No evidence for a strong resonance have been observed at the energy of 13.5 MeV in either the elastic or inelastic channels. However, if we assume that the dominant configuration for this state is $^6$He$(2^+)+\alpha$, and that the coupling to the $^6$He(g.s.)$+\alpha$ channel is relatively small, as suggested in \cite{Kravvaris2018}, then the experimentally observed spectrum can place an upper limit on the ratio between the partial width of the elastic channel to the partial width of the inelastic channels, $\Gamma_{\alpha}/\Gamma_{\alpha'} < 0.017$. This experimental information provides important constraints on the theoretical models describing the $^{10}$Be. We have further demonstrated that our data is consistent with the absence of any strong resonances in the cross section for elastic or inelastic scattering. It is clear, however, that further experiments are needed to elucidate the existence and properties of the 6$^+$ state at 13.5 MeV in $^{10}$Be.

\begin{acknowledgments} 
The authors are grateful to the cyclotron team and the Cyclotron Institute for consistently reliable operation. This work was supported by the U.S. Department of Energy, Office of Science under grant number \#DE-FG02-93ER40773, and by National Nuclear Security Administration through the Center for Excellence in Nuclear Training and University Based Research (CENTAUR) under grant number \#DE-NA0003841. Authors G.V.R. and H.J. are also supported in part by the Welch Foundation (Grant No. A-1853). Author A.V. is supported by the U.S. Department of Energy, Office of Science under grant number \#DE-SC0009883.

\end{acknowledgments}

\bibliography{10BeRefs}

\begin{thebibliography}{26}
\expandafter\ifx\csname natexlab\endcsname\relax\def\natexlab#1{#1}\fi
\expandafter\ifx\csname bibnamefont\endcsname\relax
  \def\bibnamefont#1{#1}\fi
\expandafter\ifx\csname bibfnamefont\endcsname\relax
  \def\bibfnamefont#1{#1}\fi
\expandafter\ifx\csname citenamefont\endcsname\relax
  \def\citenamefont#1{#1}\fi
\expandafter\ifx\csname url\endcsname\relax
  \def\url#1{\texttt{#1}}\fi
\expandafter\ifx\csname urlprefix\endcsname\relax\def\urlprefix{URL }\fi
\providecommand{\bibinfo}[2]{#2}
\providecommand{\eprint}[2][]{\url{#2}}

\bibitem[{\citenamefont{Okabe et~al.}(1977)\citenamefont{Okabe, Abe, and
  Tanaka}}]{Okabe1977}
\bibinfo{author}{\bibfnamefont{S.}~\bibnamefont{Okabe}},
  \bibinfo{author}{\bibfnamefont{Y.}~\bibnamefont{Abe}}, \bibnamefont{and}
  \bibinfo{author}{\bibfnamefont{H.}~\bibnamefont{Tanaka}},
  \bibinfo{journal}{Prog. Theor. Phys.} \textbf{\bibinfo{volume}{57}},
  \bibinfo{pages}{866} (\bibinfo{year}{1977}).

\bibitem[{\citenamefont{Okabe and Abe}(1979)}]{Okabe1979}
\bibinfo{author}{\bibfnamefont{S.}~\bibnamefont{Okabe}} \bibnamefont{and}
  \bibinfo{author}{\bibfnamefont{Y.}~\bibnamefont{Abe}},
  \bibinfo{journal}{Prog. Theor. Phys.} \textbf{\bibinfo{volume}{61}},
  \bibinfo{pages}{1049} (\bibinfo{year}{1979}).

\bibitem[{\citenamefont{Seya et~al.}(1981)\citenamefont{Seya, Kohno, and
  Nagata}}]{Seya1981}
\bibinfo{author}{\bibfnamefont{M.}~\bibnamefont{Seya}},
  \bibinfo{author}{\bibfnamefont{M.}~\bibnamefont{Kohno}}, \bibnamefont{and}
  \bibinfo{author}{\bibfnamefont{S.}~\bibnamefont{Nagata}},
  \bibinfo{journal}{Prog. Theor. Phys.} \textbf{\bibinfo{volume}{65}},
  \bibinfo{pages}{204} (\bibinfo{year}{1981}).

\bibitem[{\citenamefont{von Oertzen}(1997)}]{Oertzen1997}
\bibinfo{author}{\bibfnamefont{W.}~\bibnamefont{von Oertzen}},
  \bibinfo{journal}{Z. Phys. A} \textbf{\bibinfo{volume}{357}},
  \bibinfo{pages}{355} (\bibinfo{year}{1997}).

\bibitem[{\citenamefont{Kanada-En'yo et~al.}(1999)\citenamefont{Kanada-En'yo,
  Horiuchi, and Dote}}]{Kanada1999}
\bibinfo{author}{\bibfnamefont{Y.}~\bibnamefont{Kanada-En'yo}},
  \bibinfo{author}{\bibfnamefont{H.}~\bibnamefont{Horiuchi}}, \bibnamefont{and}
  \bibinfo{author}{\bibfnamefont{A.}~\bibnamefont{Dote}},
  \bibinfo{journal}{Phys. Rev. C} \textbf{\bibinfo{volume}{60}},
  \bibinfo{pages}{064304} (\bibinfo{year}{1999}).

\bibitem[{\citenamefont{Ito and Ikeda}(2014)}]{Ito2014}
\bibinfo{author}{\bibfnamefont{M.}~\bibnamefont{Ito}} \bibnamefont{and}
  \bibinfo{author}{\bibfnamefont{K.}~\bibnamefont{Ikeda}},
  \bibinfo{journal}{Rep. Prog. Phys.} \textbf{\bibinfo{volume}{77}},
  \bibinfo{pages}{096301} (\bibinfo{year}{2014}).

\bibitem[{\citenamefont{Ito}(2006)}]{Ito2006}
\bibinfo{author}{\bibfnamefont{M.}~\bibnamefont{Ito}},
  \bibinfo{journal}{Physics Letters, Section B: Nuclear, Elementary Particle
  and High-Energy Physics} \textbf{\bibinfo{volume}{636}}, \bibinfo{pages}{293}
  (\bibinfo{year}{2006}), ISSN \bibinfo{issn}{03702693}.

\bibitem[{\citenamefont{Dot{\'{e}} et~al.}(1997)\citenamefont{Dot{\'{e}},
  Horiuchi, and Kanada-En'yo}}]{Dote1997}
\bibinfo{author}{\bibfnamefont{A.}~\bibnamefont{Dot{\'{e}}}},
  \bibinfo{author}{\bibfnamefont{H.}~\bibnamefont{Horiuchi}}, \bibnamefont{and}
  \bibinfo{author}{\bibfnamefont{Y.}~\bibnamefont{Kanada-En'yo}},
  \bibinfo{journal}{Physical Review C} \textbf{\bibinfo{volume}{56}},
  \bibinfo{pages}{1844} (\bibinfo{year}{1997}), ISSN \bibinfo{issn}{0556-2813},
  \urlprefix\url{https://link.aps.org/doi/10.1103/PhysRevC.56.1844}.

\bibitem[{\citenamefont{{Abe Y., Hiura J.}}(1973)}]{Abe1973}
\bibinfo{author}{\bibfnamefont{T.~H.} \bibnamefont{{Abe Y., Hiura J.}}},
  \bibinfo{journal}{Prog. Theor. Phys.} \textbf{\bibinfo{volume}{49}},
  \bibinfo{pages}{800} (\bibinfo{year}{1973}).

\bibitem[{\citenamefont{Nishioka}(1984)}]{Nishioka1984}
\bibinfo{author}{\bibfnamefont{H.}~\bibnamefont{Nishioka}},
  \bibinfo{journal}{Journal of Physics G: Nuclear Physics}
  \textbf{\bibinfo{volume}{10}}, \bibinfo{pages}{1713} (\bibinfo{year}{1984}),
  \urlprefix\url{http://stacks.iop.org/0305-4616/10/i=12/a=010}.

\bibitem[{\citenamefont{Hamada et~al.}(1994)\citenamefont{Hamada, Yasue,
  Kubono, Tanaka, and Peterson}}]{Hamada1994}
\bibinfo{author}{\bibfnamefont{S.}~\bibnamefont{Hamada}},
  \bibinfo{author}{\bibfnamefont{M.}~\bibnamefont{Yasue}},
  \bibinfo{author}{\bibfnamefont{S.}~\bibnamefont{Kubono}},
  \bibinfo{author}{\bibfnamefont{M.~H.} \bibnamefont{Tanaka}},
  \bibnamefont{and} \bibinfo{author}{\bibfnamefont{R.~J.}
  \bibnamefont{Peterson}}, \bibinfo{journal}{Phys. Rev. C}
  \textbf{\bibinfo{volume}{49}}, \bibinfo{pages}{3192} (\bibinfo{year}{1994}),
  \urlprefix\url{https://link.aps.org/doi/10.1103/PhysRevC.49.3192}.

\bibitem[{\citenamefont{Curtis et~al.}(2001)\citenamefont{Curtis, Caussyn,
  Fletcher, Mar{\'{e}}chal, Fay, and Robson}}]{Curtis2001}
\bibinfo{author}{\bibfnamefont{N.}~\bibnamefont{Curtis}},
  \bibinfo{author}{\bibfnamefont{D.~D.} \bibnamefont{Caussyn}},
  \bibinfo{author}{\bibfnamefont{N.~R.} \bibnamefont{Fletcher}},
  \bibinfo{author}{\bibfnamefont{F.}~\bibnamefont{Mar{\'{e}}chal}},
  \bibinfo{author}{\bibfnamefont{N.}~\bibnamefont{Fay}}, \bibnamefont{and}
  \bibinfo{author}{\bibfnamefont{D.}~\bibnamefont{Robson}},
  \bibinfo{journal}{Physical Review C} \textbf{\bibinfo{volume}{64}},
  \bibinfo{pages}{044604} (\bibinfo{year}{2001}), ISSN
  \bibinfo{issn}{0556-2813},
  \urlprefix\url{https://link.aps.org/doi/10.1103/PhysRevC.64.044604}.

\bibitem[{\citenamefont{Milin et~al.}(2005)\citenamefont{Milin, Zadro,
  Cherubini, Davinson, {Di Pietro}, Figuera, Miljani{\'{c}}, Musumarra, Ninane,
  Ostrowski et~al.}}]{Milin2005}
\bibinfo{author}{\bibfnamefont{M.}~\bibnamefont{Milin}},
  \bibinfo{author}{\bibfnamefont{M.}~\bibnamefont{Zadro}},
  \bibinfo{author}{\bibfnamefont{S.}~\bibnamefont{Cherubini}},
  \bibinfo{author}{\bibfnamefont{T.}~\bibnamefont{Davinson}},
  \bibinfo{author}{\bibfnamefont{A.}~\bibnamefont{{Di Pietro}}},
  \bibinfo{author}{\bibfnamefont{P.}~\bibnamefont{Figuera}},
  \bibinfo{author}{\bibfnamefont{Ä.}~\bibnamefont{Miljani{\'{c}}}},
  \bibinfo{author}{\bibfnamefont{A.}~\bibnamefont{Musumarra}},
  \bibinfo{author}{\bibfnamefont{A.}~\bibnamefont{Ninane}},
  \bibinfo{author}{\bibfnamefont{A.}~\bibnamefont{Ostrowski}},
  \bibnamefont{et~al.}, \bibinfo{journal}{Nuclear Physics A}
  \textbf{\bibinfo{volume}{753}}, \bibinfo{pages}{263} (\bibinfo{year}{2005}),
  ISSN \bibinfo{issn}{03759474},
  \urlprefix\url{http://www.sciencedirect.com/science/article/pii/S0375947405003398
  http://linkinghub.elsevier.com/retrieve/pii/S0375947405003398}.

\bibitem[{\citenamefont{Freer et~al.}(2006)\citenamefont{Freer, Casarejos,
  Achouri, Angulo, Ashwood, Curtis, Demaret, Harlin, Laurent, Milin
  et~al.}}]{Freer2006}
\bibinfo{author}{\bibfnamefont{M.}~\bibnamefont{Freer}},
  \bibinfo{author}{\bibfnamefont{E.}~\bibnamefont{Casarejos}},
  \bibinfo{author}{\bibfnamefont{L.}~\bibnamefont{Achouri}},
  \bibinfo{author}{\bibfnamefont{C.}~\bibnamefont{Angulo}},
  \bibinfo{author}{\bibfnamefont{N.~I.} \bibnamefont{Ashwood}},
  \bibinfo{author}{\bibfnamefont{N.}~\bibnamefont{Curtis}},
  \bibinfo{author}{\bibfnamefont{P.}~\bibnamefont{Demaret}},
  \bibinfo{author}{\bibfnamefont{C.}~\bibnamefont{Harlin}},
  \bibinfo{author}{\bibfnamefont{B.}~\bibnamefont{Laurent}},
  \bibinfo{author}{\bibfnamefont{M.}~\bibnamefont{Milin}},
  \bibnamefont{et~al.}, \bibinfo{journal}{Phys. Rev. Lett.}
  \textbf{\bibinfo{volume}{96}}, \bibinfo{pages}{042501}
  (\bibinfo{year}{2006}),
  \urlprefix\url{https://link.aps.org/doi/10.1103/PhysRevLett.96.042501}.

\bibitem[{\citenamefont{Wolsky et~al.}(2010)\citenamefont{Wolsky, Gnilozub,
  Kurgalin, and Tchuvil'sky}}]{Wolsky2010}
\bibinfo{author}{\bibfnamefont{R.}~\bibnamefont{Wolsky}},
  \bibinfo{author}{\bibfnamefont{I.~A.} \bibnamefont{Gnilozub}},
  \bibinfo{author}{\bibfnamefont{S.~D.} \bibnamefont{Kurgalin}},
  \bibnamefont{and} \bibinfo{author}{\bibfnamefont{Y.~M.}
  \bibnamefont{Tchuvil'sky}}, \bibinfo{journal}{Physics of Atomic Nuclei}
  \textbf{\bibinfo{volume}{73}}, \bibinfo{pages}{1405} (\bibinfo{year}{2010}),
  ISSN \bibinfo{issn}{1562-692X},
  \urlprefix\url{https://doi.org/10.1134/S1063778810080144}.

\bibitem[{\citenamefont{Rogachev et~al.}(2014)\citenamefont{Rogachev, Avila,
  Kuchera, Baby, Belarge, Blackmon, Goldberg, Johnson, Kemper, Koshchiy
  et~al.}}]{Rogachev2014}
\bibinfo{author}{\bibfnamefont{G.~V.} \bibnamefont{Rogachev}},
  \bibinfo{author}{\bibfnamefont{M.~L.} \bibnamefont{Avila}},
  \bibinfo{author}{\bibfnamefont{A.~N.} \bibnamefont{Kuchera}},
  \bibinfo{author}{\bibfnamefont{L.~T.} \bibnamefont{Baby}},
  \bibinfo{author}{\bibfnamefont{J.}~\bibnamefont{Belarge}},
  \bibinfo{author}{\bibfnamefont{J.}~\bibnamefont{Blackmon}},
  \bibinfo{author}{\bibfnamefont{V.~Z.} \bibnamefont{Goldberg}},
  \bibinfo{author}{\bibfnamefont{E.~D.} \bibnamefont{Johnson}},
  \bibinfo{author}{\bibfnamefont{K.~W.} \bibnamefont{Kemper}},
  \bibinfo{author}{\bibfnamefont{E.}~\bibnamefont{Koshchiy}},
  \bibnamefont{et~al.}, \bibinfo{journal}{Journal of Physics: Conference
  Series} \textbf{\bibinfo{volume}{569}}, \bibinfo{pages}{012004}
  (\bibinfo{year}{2014}),
  \urlprefix\url{http://stacks.iop.org/1742-6596/569/i=1/a=012004}.

\bibitem[{\citenamefont{Kuchera}(2013)}]{Kuchera2013}
\bibinfo{author}{\bibfnamefont{A.~N.} \bibnamefont{Kuchera}}, Ph.D. thesis,
  \bibinfo{school}{Florida State University} (\bibinfo{year}{2013}),
  \urlprefix\url{http://diginole.lib.fsu.edu/cgi/viewcontent.cgi?article=7813{\&}context=etd}.

\bibitem[{\citenamefont{Dell'Aquila et~al.}(2016)\citenamefont{Dell'Aquila,
  Lombardo, Acosta, Andolina, Auditore, Cardella, Chatterjiee, {De Filippo},
  Francalanza, Gnoffo et~al.}}]{DellAquila2016}
\bibinfo{author}{\bibfnamefont{D.}~\bibnamefont{Dell'Aquila}},
  \bibinfo{author}{\bibfnamefont{I.}~\bibnamefont{Lombardo}},
  \bibinfo{author}{\bibfnamefont{L.}~\bibnamefont{Acosta}},
  \bibinfo{author}{\bibfnamefont{R.}~\bibnamefont{Andolina}},
  \bibinfo{author}{\bibfnamefont{L.}~\bibnamefont{Auditore}},
  \bibinfo{author}{\bibfnamefont{G.}~\bibnamefont{Cardella}},
  \bibinfo{author}{\bibfnamefont{M.~B.} \bibnamefont{Chatterjiee}},
  \bibinfo{author}{\bibfnamefont{E.}~\bibnamefont{{De Filippo}}},
  \bibinfo{author}{\bibfnamefont{L.}~\bibnamefont{Francalanza}},
  \bibinfo{author}{\bibfnamefont{B.}~\bibnamefont{Gnoffo}},
  \bibnamefont{et~al.}, \bibinfo{journal}{Physical Review C}
  \textbf{\bibinfo{volume}{93}}, \bibinfo{pages}{024611}
  (\bibinfo{year}{2016}), ISSN \bibinfo{issn}{24699993}.

\bibitem[{\citenamefont{Jiang et~al.}(2017)\citenamefont{Jiang, Ye, Li, Lin,
  Li, Ge, Lou, Jiang, Li, Tian et~al.}}]{Jiang2017}
\bibinfo{author}{\bibfnamefont{W.}~\bibnamefont{Jiang}},
  \bibinfo{author}{\bibfnamefont{Y.}~\bibnamefont{Ye}},
  \bibinfo{author}{\bibfnamefont{Z.}~\bibnamefont{Li}},
  \bibinfo{author}{\bibfnamefont{C.}~\bibnamefont{Lin}},
  \bibinfo{author}{\bibfnamefont{Q.}~\bibnamefont{Li}},
  \bibinfo{author}{\bibfnamefont{Y.}~\bibnamefont{Ge}},
  \bibinfo{author}{\bibfnamefont{J.}~\bibnamefont{Lou}},
  \bibinfo{author}{\bibfnamefont{D.}~\bibnamefont{Jiang}},
  \bibinfo{author}{\bibfnamefont{J.}~\bibnamefont{Li}},
  \bibinfo{author}{\bibfnamefont{Z.}~\bibnamefont{Tian}}, \bibnamefont{et~al.},
  \bibinfo{journal}{Science China Physics, Mechanics {\&} Astronomy}
  \textbf{\bibinfo{volume}{60}}, \bibinfo{pages}{062011}
  (\bibinfo{year}{2017}), ISSN \bibinfo{issn}{1674-7348},
  \urlprefix\url{http://link.springer.com/10.1007/s11433-017-9023-x}.

\bibitem[{\citenamefont{Tribble et~al.}(1989)\citenamefont{Tribble, Burch, and
  Gagliardi}}]{MARS}
\bibinfo{author}{\bibfnamefont{R.~E.} \bibnamefont{Tribble}},
  \bibinfo{author}{\bibfnamefont{R.~H.} \bibnamefont{Burch}}, \bibnamefont{and}
  \bibinfo{author}{\bibfnamefont{C.~A.} \bibnamefont{Gagliardi}},
  \bibinfo{journal}{Nuclear Inst. and Methods in Physics Research, A}
  \textbf{\bibinfo{volume}{285}}, \bibinfo{pages}{441} (\bibinfo{year}{1989}),
  ISSN \bibinfo{issn}{01689002}.

\bibitem[{mic()}]{micron}
\emph{\bibinfo{title}{Micron msq25-1000}},
  \urlprefix\url{http://www.micronsemiconductor.co.uk/product/msq25/}.

\bibitem[{\citenamefont{Johnson}(2008)}]{Johnson2008}
\bibinfo{author}{\bibfnamefont{E.~D.} \bibnamefont{Johnson}}, Ph.D. thesis,
  \bibinfo{school}{Florida State University} (\bibinfo{year}{2008}).

\bibitem[{\citenamefont{Azuma et~al.}(2010)\citenamefont{Azuma, Uberseder,
  Simpson, Brune, Costantini, de~Boer, G\"orres, Heil, LeBlanc, Ugalde
  et~al.}}]{Azuma2010AZURE}
\bibinfo{author}{\bibfnamefont{R.~E.} \bibnamefont{Azuma}},
  \bibinfo{author}{\bibfnamefont{E.}~\bibnamefont{Uberseder}},
  \bibinfo{author}{\bibfnamefont{E.~C.} \bibnamefont{Simpson}},
  \bibinfo{author}{\bibfnamefont{C.~R.} \bibnamefont{Brune}},
  \bibinfo{author}{\bibfnamefont{H.}~\bibnamefont{Costantini}},
  \bibinfo{author}{\bibfnamefont{R.~J.} \bibnamefont{de~Boer}},
  \bibinfo{author}{\bibfnamefont{J.}~\bibnamefont{G\"orres}},
  \bibinfo{author}{\bibfnamefont{M.}~\bibnamefont{Heil}},
  \bibinfo{author}{\bibfnamefont{P.~J.} \bibnamefont{LeBlanc}},
  \bibinfo{author}{\bibfnamefont{C.}~\bibnamefont{Ugalde}},
  \bibnamefont{et~al.}, \bibinfo{journal}{Phys. Rev. C}
  \textbf{\bibinfo{volume}{81}}, \bibinfo{pages}{045805}
  (\bibinfo{year}{2010}),
  \urlprefix\url{https://link.aps.org/doi/10.1103/PhysRevC.81.045805}.

\bibitem[{\citenamefont{Kravvaris and Volya}(2018)}]{Kravvaris2018}
\bibinfo{author}{\bibfnamefont{K.}~\bibnamefont{Kravvaris}} \bibnamefont{and}
  \bibinfo{author}{\bibfnamefont{A.}~\bibnamefont{Volya}}, in
  \emph{\bibinfo{booktitle}{AIP Conference Proceedings}}
  (\bibinfo{year}{2018}), vol. \bibinfo{volume}{2038}, p.
  \bibinfo{pages}{020026}, ISBN \bibinfo{isbn}{9780735417649}, ISSN
  \bibinfo{issn}{15517616},
  \urlprefix\url{http://aip.scitation.org/doi/abs/10.1063/1.5078845}.

\bibitem[{\citenamefont{Kravvaris and Volya}(2019)}]{Kravvaris2019}
\bibinfo{author}{\bibfnamefont{K.}~\bibnamefont{Kravvaris}} \bibnamefont{and}
  \bibinfo{author}{\bibfnamefont{A.}~\bibnamefont{Volya}},
  \bibinfo{journal}{Phys. Rev. C} \textbf{\bibinfo{volume}{100}},
  \bibinfo{pages}{034321} (\bibinfo{year}{2019}).

\bibitem[{\citenamefont{Suzuki et~al.}(2013)\citenamefont{Suzuki, Shore,
  Mittig, Kolata, Bazin, Ford, Ahn, Becchetti, Beceiro~Novo, Ben~Ali
  et~al.}}]{Suzuki2013}
\bibinfo{author}{\bibfnamefont{D.}~\bibnamefont{Suzuki}},
  \bibinfo{author}{\bibfnamefont{A.}~\bibnamefont{Shore}},
  \bibinfo{author}{\bibfnamefont{W.}~\bibnamefont{Mittig}},
  \bibinfo{author}{\bibfnamefont{J.~J.} \bibnamefont{Kolata}},
  \bibinfo{author}{\bibfnamefont{D.}~\bibnamefont{Bazin}},
  \bibinfo{author}{\bibfnamefont{M.}~\bibnamefont{Ford}},
  \bibinfo{author}{\bibfnamefont{T.}~\bibnamefont{Ahn}},
  \bibinfo{author}{\bibfnamefont{F.~D.} \bibnamefont{Becchetti}},
  \bibinfo{author}{\bibfnamefont{S.}~\bibnamefont{Beceiro~Novo}},
  \bibinfo{author}{\bibfnamefont{D.}~\bibnamefont{Ben~Ali}},
  \bibnamefont{et~al.}, \bibinfo{journal}{Phys. Rev. C}
  \textbf{\bibinfo{volume}{87}}, \bibinfo{pages}{054301}
  (\bibinfo{year}{2013}),
  \urlprefix\url{https://link.aps.org/doi/10.1103/PhysRevC.87.054301}.

\end{thebibliography}

\end{document}